\documentclass[%
 reprint,
 amsmath,amssymb,
 aps,
pra,
floatfix,
]{revtex4-2}

\usepackage{graphicx}
\usepackage{dcolumn}
\usepackage{bm}

\usepackage{braket}
\usepackage{bbm}
\usepackage[english]{babel}
\usepackage{natbib} 
\usepackage{booktabs} 
\usepackage{array}    

\begin{document}
\selectlanguage{english}
\title{Three wave mixing vacuum squeezing generation in a SNAIL-based Traveling-Wave Parametric Amplifier with alternated flux polarity}

\author{Isita Chatterjee$^{1,2}$}
\author{Pegah Darvehi$^{1}$}
\author{Antonio Orsi$^{1,2}$}
\author{Anna Levochkina$^{1,2}$}
\author{Pasquale Mastrovito$^{2,1}$}
\author{Gwenael Le Gal$^{3}$}
\thanks{Currently at CEA, Leti, Grenoble, France}
\author{Arpit Ranadive$^{3}$}
\thanks{Currently at Google Quantum AI, Santa Barbara, CA, USA}
\author{Giulio Cappelli$^{3}$}
\thanks{Currently at Fondazione Bruno Kessler (FBK), Trento, Italy}
\author{Alberto Porzio$^{4}$}
\author{Francesco Tafuri$^{2}$}
\author{Davide Massarotti$^{5, 1}$}
\author{Martina Esposito$^{1}$}
\email{Corresponding author: martina.esposito@cnr.it}

\affiliation{
$^1$ CNR-SPIN Complesso di Monte S. Angelo, 80126 Napoli, Italy \\
$^2$ Dipartimento di Fisica “Ettore Pancini”, Università di Napoli “Federico II,” Monte S. Angelo, I-80126 Napoli, Italy \\
$^3$ Université Grenoble Alpes, CNRS, Grenoble INP, Institut Néel, 38000 Grenoble, France\\
$^4$ Dipartimento di Ingegneria Civile e Meccanica (DICEM), Università di Cassino e Lazio Meridionale, Cassino, Italy \\
$^5$ Dipartimento di Ingegneria Elettrica e delle Tecnologie dell’Informazione, Università degli Studi di Napoli Federico II, via Claudio, I-80125 Napoli, Italy}

\begin{abstract}
Recent demonstrations of squeezing generation using Traveling Wave Parametric Amplifiers (TWPAs) have opened the way for the application of broadband microwave squeezing in quantum sensing, quantum-enhanced detection, and continuous-variable quantum information. 
Here we demonstrate vacuum squeezing generation via residual three-wave mixing (3WM) in a Josephson TWPA based on superconducting nonlinear asymmetric inductive elements (SNAILs) with alternated magnetic flux polarity. 
By investigating competition between four-wave mixing (4WM) and 3WM nonlinearities, we prove that vacuum squeezing generation via residual 3WM is possible when a careful choice of the operating flux point is adopted. Our study provides valuable insights on the impact of competing nonlinearities on TWPA squeezers, potentially extending the range of applications in the framework of microwave photonics.

\end{abstract}
\maketitle

\section{Introduction} 
Squeezed states of the electromagnetic radiation play a central role in quantum technologies, since they allow noise reduction below the standard quantum limit (SQL) by redistributing quantum fluctuations among conjugate field quadratures \cite{fabre_modes_2020}. While squeezing was initially extensively studied in the optical frequency domain, significant progress over the past decade in the field of superconducting quantum circuits has driven intense research efforts for squeezing generation in the microwave regime \cite{gu_microwave_2017, casariego_propagating_2023}. Applications of microwave squeezing generation include, for example, quantum-enhanced detection \cite{ malnou_squeezed_2019,backes2021quantum,fasolo_josephson_2021}, continuous-variable quantum information \cite{feng_microwave_2025}, quantum communication \cite{pogorzalek_secure_2019}, and entanglement transfer protocols \cite{agusti_long-distance_2022,andres-juanes_entangling_2025}.

Josephson-junction-based parametric amplifiers provide controllable nonlinearities that enable the generation of microwave squeezed states \cite{Aumentado20}.  
Early implementations of microwave squeezing relied on resonator-based Josephson parametric amplifiers (JPAs), showing remarkable squeezing performance with bandwidth typically limited to the range of 10-100 MHz \cite{boutin_effect_2017,malnouOptimalOperationJosephson2018,bienfaitMagneticResonanceSqueezed2017}.

To address the demand of broadband microwave squeezing generation, Josephson Traveling Wave Parametric Amplifiers (JTWPAs) have recently emerged as sources of non-classical microwave radiation, both in 3WM \cite{Perelshtein22, alocco_programmable_2025} and 4WM \cite{qiu2023broadband, Esposito22} regimes. However, the impact of competition between 3WM and 4WM nonlinearities on squeezing generation in TWPA devices has not yet been investigated.

Here, we report an experimental study of vacuum squeezing generation in a prototypical flux-tunable JTWPA based on superconducting nonlinear asymmetric inductive element (SNAIL) \cite{frattini_3-wave_2017} unit cells with alternated magnetic flux polarity. Such a device was first introduced in the literature as a 4WM amplifier \cite{Ranadive22} and 4WM squeezer \cite{Esposito22}. Subsequent investigations have shown that typical Josephson junctions’ fabrication imperfections can activate residual 3WM, allowing second harmonic generation \cite{levochkina_investigating_2024} and dynamic phase matching amplification \cite{ranadive_travelling-wave_2025}. Here, we demonstrate for the first time that residual 3WM can also be used for squeezing generation, provided that competing 4WM nonlinear processes are mitigated via an appropriate choice of the magnetic flux operating point.

We study the influence of competing 4WM processes on squeezing generation, by performing 3WM phase-dependent gain, single-mode squeezing, and two-mode squeezing experiments for two significant flux operating points, eventually providing an experimental procedure to identify optimal flux bias points for 3WM vacuum squeezing performance. 
The demonstrated 3WM squeezing regime has the advantage of a large separation in frequency between the pump and the squeezed photons, allowing for straightforward filtering out of the pump and thus facilitating applications, for example, in the context of superconducting quantum networks. 

\section{Flux tunability preliminary characterization}

\begin{figure}[ht]
    \centering
    \includegraphics[width=0.5\textwidth]{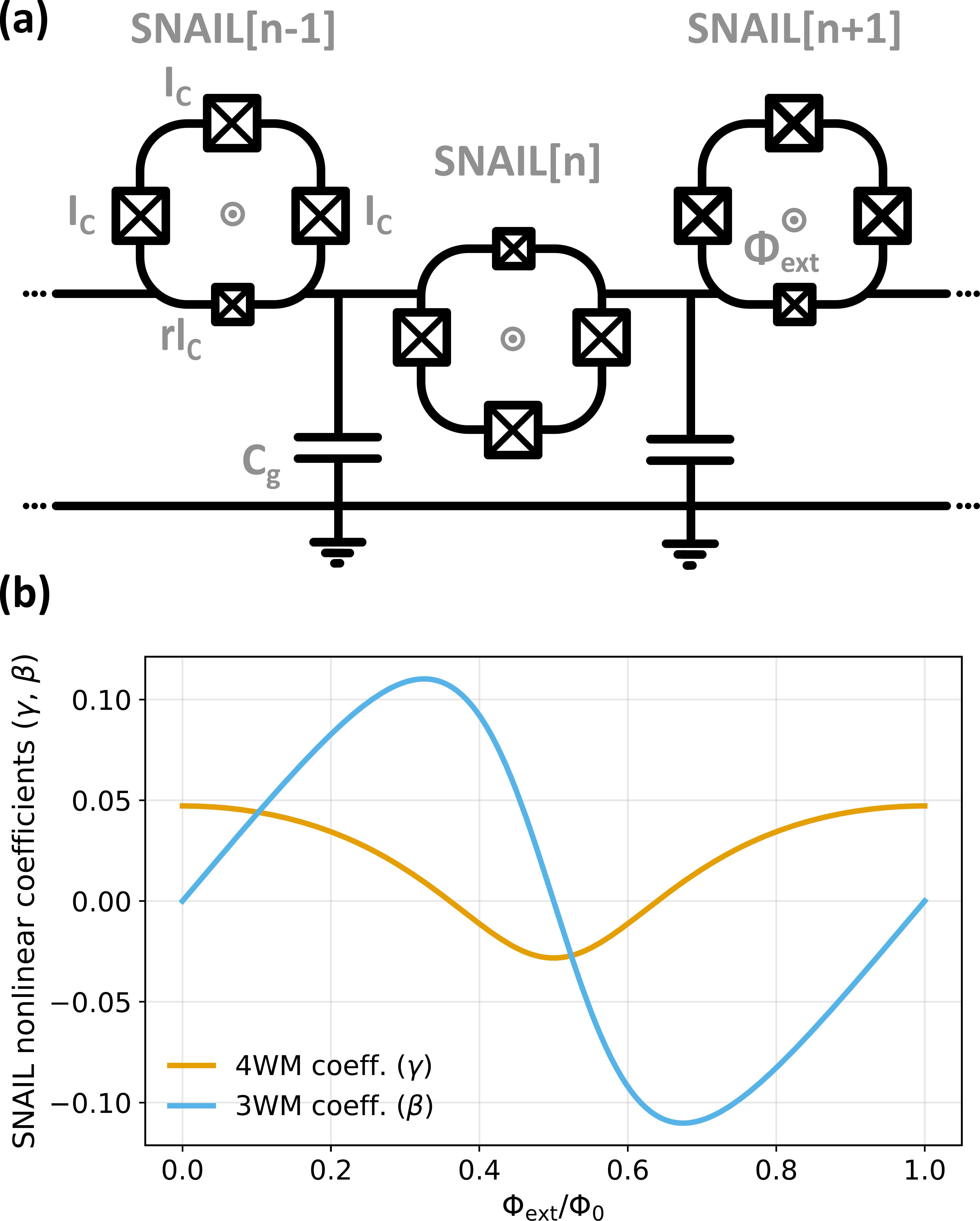}
    \caption{\textbf{Device sketch and SNAIL flux tunability.} \textbf{(a)} Circuit schematic of three neighboring unit cells in a SNAIL-based JTWPA device with alternated magnetic flux polarity. \textbf{(b) }3WM coefficient $\beta$ and 4WM coefficient $\gamma$ for one SNAIL as a function of external magnetic flux.} 
    \label{3wm}
\end{figure}

The adopted device is a SNAIL-based JTWPA with alternated flux polarity throughout the transmission line, as shown in the sketch in Fig.  \ref{3wm}(a). Each unit cell is composed of a SNAIL element and a capacitance to ground, $C_g$. Each SNAIL is a superconducting loop with three identical Josephson junctions with critical current $I_C$ in one arm, and one smaller Josephson junction with critical current $r I_C$ in the other arm. 
By using Taylor expansion, the current phase relation of a SNAIL (see Appendix \ref{beta_gamma}) can be approximated as
\begin{equation}
    \frac{I(\phi^*+\phi)}{\tilde{\alpha} I_c}  \approx \phi-\beta \phi^2-\gamma\phi^3,
    \label{curr_phi}
\end{equation}
where $\phi^*$ is such that $I(\phi^*)=0$, and the coefficients are defined as follows
\begin{equation}
\tilde{\alpha} = r\cos(\phi^{*})
+ \frac{1}{3}\cos\!\left(\frac{\phi^{*} - \phi_{\mathrm{ext}}}{3}\right) \, ,
\label{alpha}
\end{equation}
\begin{equation}
\beta = \frac{1}{2}
\left[
    r \sin\phi^{*}
    + \frac{1}{9}\sin\!\left(\frac{\phi^{*}-\phi_{\mathrm{ext}}}{3}\right)\right]/\tilde{\alpha} \, ,
\label{beta}
\end{equation}
\begin{equation}
\gamma = \frac{1}{6}
\left[
    r \cos\phi^{*}
    + \frac{1}{27}\cos\!\left(\frac{\phi^{*}-\phi_{\mathrm{ext}}}{3}\right)\right]/\tilde{\alpha} \, .
\label{gamma}
\end{equation}
The coefficients $\beta$ and $\gamma$ are the 3WM and 4WM nonlinear coefficients respectively, while $\phi_{\text{ext}}$=$2\pi \Phi_{\text{ext}}/\Phi_0$ is the reduced external magnetic flux, with $\Phi_0$=$h/(2e)$ the magnetic flux quantum. 
Fig. \ref{3wm}(b) reports the flux-dependent behavior of $\beta$ and $\gamma$ for one SNAIL, considering the design parameter of the adopted device.
Since the $\beta$ coefficient is an odd function of the external flux, the alternated magnetic flux polarity configuration should ideally provide an overall suppression of 3WM processes at the TWPA output for wavelengths much larger than the unit cell size. However, as demonstrated in our previous study \cite{levochkina_investigating_2024}, Josephson junctions’ fabrication imperfections can lead to a residual 3WM nonlinearity, which for example has been recently implemented for dynamic phase matching amplification \cite{ranadive_travelling-wave_2025}.  In the following, we experimentally demonstrate that such residual 3WM nonlinearity can be employed for squeezing generation.

To identify suitable flux operating points for squeezing generation with residual 3WM, we perform a flux-dependent measurement of the generated idler power spectral density (PSD) with a spectrum analyzer, both in 3WM and 4WM configurations. The experimental results reporting the measured idler power referred at the TWPA output are shown in Fig. \ref{fig:flux_sweet_spot} (b).

The measurement is performed using the setup described in Appendix~\ref{exp_setup}, where the external magnetic flux is controlled by injecting DC current in a superconducting coil underneath the device.
By applying a pump tone at frequency $f_p$ along with a signal tone at frequency $f_s$, we measure the power spectral density at the idler frequency $f_i = 2f_p - f_s$ (and $f_i = f_p - f_s$) for the 4WM (and 3WM) case as a function of the external flux. For the 3WM configuration, the signal is applied at $(f_p/2 - \Delta)$, while for the 4WM case the signal is applied at $(f_p - \Delta)$, where $\Delta$ indicates the frequency detuning. The pump frequency is kept the same for both cases. A sketch of the frequency configuration for 3WM and 4WM experiments is shown in Fig. \ref{fig:flux_sweet_spot} (a). 
\begin{figure}[htbp]
    \includegraphics[width=1\linewidth]{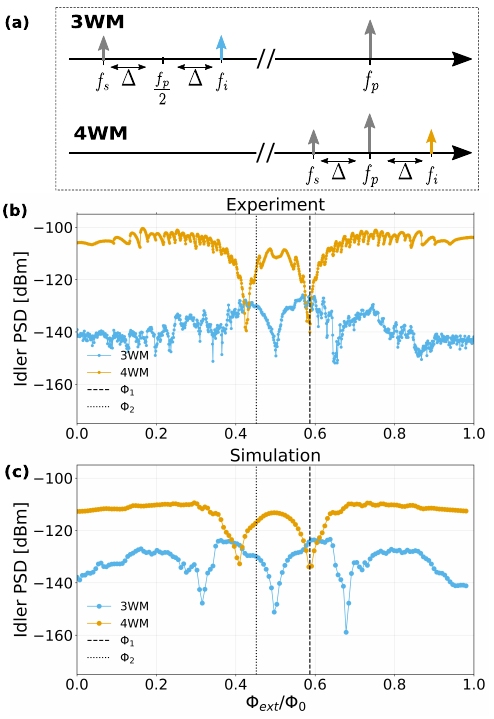}
    \caption{\textbf{Idler generation in 3WM and 4WM.} \textbf{(a)} Sketch of the frequency configuration for 3WM and 4WM experiments where $\Delta = 31\,\mathrm{MHz}$. \textbf{(b)}  Experimental results for 3WM and 4WM idler measured power referred at TWPA output vs flux;  pump frequency $f_p = 7.705\ \mathrm{GHz}$, pump power at device input $P_p = -87\ \mathrm{dBm}$, signal power at device input $P_s = -120\,\mathrm{dBm}$. \textbf{(c)} Results of WRSpice numerical simulations (see Appendix \ref{simulation}). The vertical dashed and dotted lines indicate the selected flux points.}
    \label{fig:flux_sweet_spot}
\end{figure}
The experimental results in Fig. \ref{fig:flux_sweet_spot} (b) are in qualitative agreement with WRSPICE transient simulations reported in Fig. \ref{fig:flux_sweet_spot} (c). The numerical simulations consider device parameters estimated in experimental characterizations \cite{Ranadive22}, including also the typical Josephson junctions’ fabrication imperfections (see Appendix \ref{simulation}).

Based on the above characterization of the flux dependence of the nonlinearity, we identify two significant flux working points, both corresponding to about maximum values for the 3WM idler PSD ($\sim -130 \,  \mathrm{dBm}$ at TWPA output), but which significantly differ from the point of view of the 4WM idler PSD ($\sim 12 \,  \mathrm{dB}$ difference). 
The first identified flux point, $\Phi_1=0.59 \, \Phi_0$, falls into a region of maximum 3WM idler generation and minimum 4WM idler generation. 
In contrast, the second identified flux point, $\Phi_2=0.45 \, \Phi_0$, does not minimize the 4WM idler measured PSD.
The two flux points are deliberately chosen to investigate the role of the combination of 3WM and 4WM nonlinearities on the squeezing generation in the regime of residual 3WM at the selected signal frequency. 
In the following, we report a comparative analysis of experimental results for phase-sensitive gain measurements, single-mode and two-mode squeezing experiments for these two selected working points. 
\section{Gain and squeezing experiments}
The experimental setup for gain and squeezing experiments is described in the Appendix \ref{exp_setup}.

\begin{figure*}[htbp]
    \centering
    \includegraphics[width=0.8\linewidth]{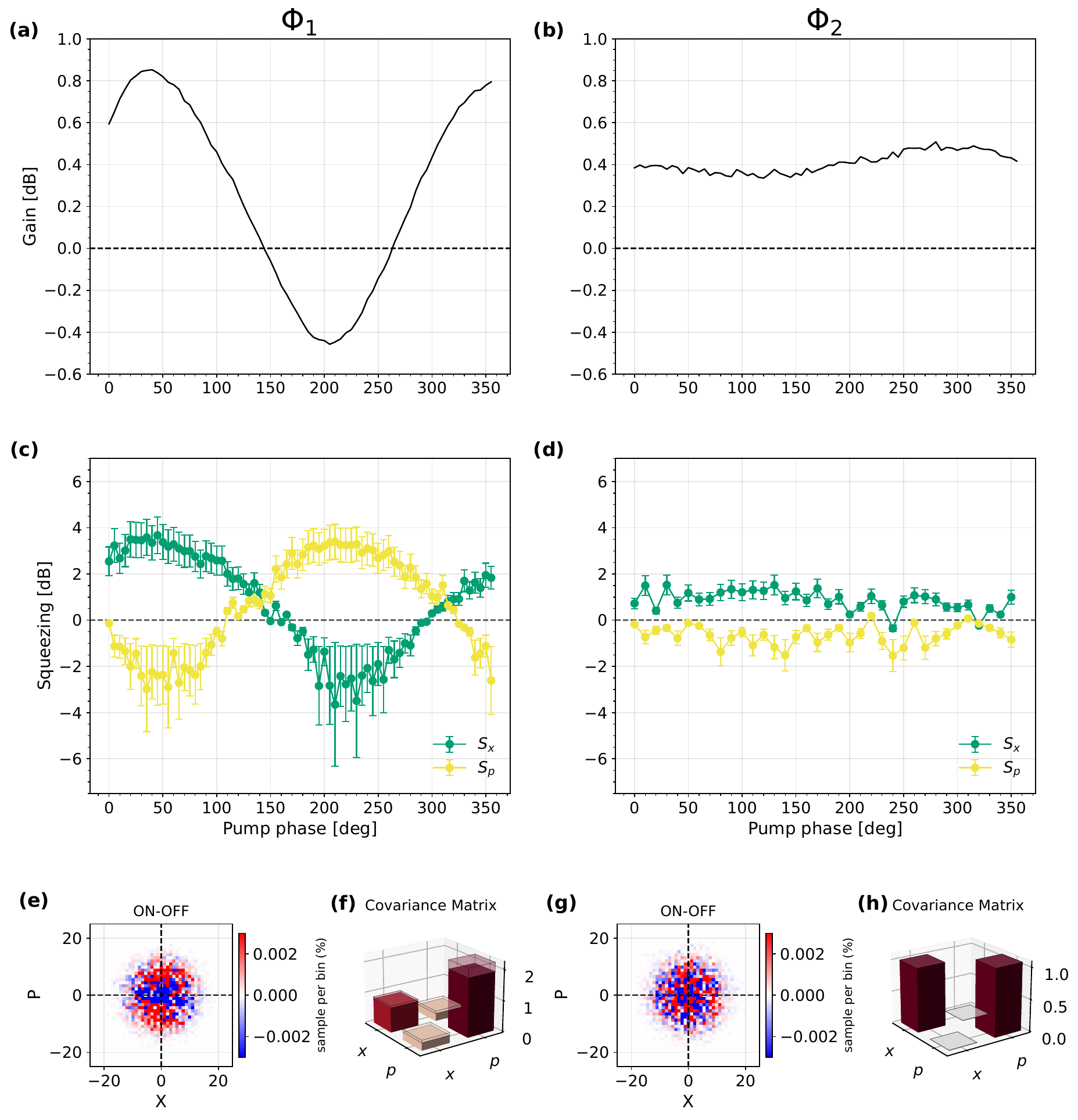}
    \caption{\textbf{Degenerate gain and single mode squeezing for $\Phi_1$ (left) and $\Phi_2$ (right).} \textbf{(a-b)} 3WM degenerate gain vs pump phase, pump frequency $f_p = 7.705~\mathrm{GHz}$, pump power at TWPA input $ P_p = -84\,\mathrm{dBm}$ for $\Phi_1$ and $ P_p = -91\,\mathrm{dBm}$ for $\Phi_2$. \textbf{(c-d)} 3WM single-mode squeezing values, $S_{x}$ and $S_{p}$, along the $x$ and $p$ quadratures as a function of pump phase; number of repeated acquisitions for each phase, \( N_{\mathrm{rep}} = 3 \times 10^{6} \) for $\Phi_1$ and \( N_{\mathrm{rep}} = 1 \times 10^{6} \) for $\Phi_2$; integration time for each acquisition, $10 \, \mu\text{s}$. \textbf{(e, g)} Examples of single-mode differential (pump ON - pump OFF) phase space histogram plots for $N_{\mathrm{rep}} = 8 \times 10^{6}$ repeated acquisitions and pump phase $220^\circ$ and corresponding inferred covariance matrix with uncertainty indicated by shaded regions \textbf{(f, h)}.}
    \label{fig:SMS_vs_pump_phase}
\end{figure*}

We first investigate the activation of residual 3WM parametric amplification by measuring the signal gain as a function of the pump phase in the degenerate configuration ($f_s = f_i = f_p/2$) for both the selected operating flux points.
Specifically, we send at TWPA input a CW pump tone at frequency $f_p$ along with a much weaker CW signal tone, locked in phase to the pump and measure the gain of the outgoing signal as a function of the input pump phase. 
In Fig. \ref{fig:SMS_vs_pump_phase} (a) and (b) we report the experimental measure of phase-dependent 3WM gain for the two operating flux points for pump frequency  $f_p = 7.705~\mathrm{GHz}$, pump power $ P_p = -84\,\mathrm{dBm}$ for $\Phi_1$ and $ P_p = -91\,\mathrm{dBm}$ for $\Phi_2$ and signal power $ P_s = -120\,\mathrm{dBm}$ for both fluxes at the device input. 

We observe significantly different results for the two flux points. For $\Phi_1$, in which 4WM idler generation is minimized, a phase-sensitive gain behavior is observed, as reported in Fig. \ref{fig:SMS_vs_pump_phase} (a), showing a clear periodic amplification and de-amplification as expected for 3WM degenerate gain, proving the activation of an amplification process due to the residual 3WM nonlinearity. On the contrary, for the $\Phi_2$ operating point, which has similar 3WM idler PSD as $\Phi_1$ but 4WM idler PSD not minimized, the periodic amplification and de-amplification are not observed, as shown in Fig. \ref{fig:SMS_vs_pump_phase} (b), suggesting a degradation of the degenerate 3WM gain due to competing 4WM processes.

To explore the role of competing nonlinearities specifically on the squeezing performance, we finally perform single-mode squeezing (SMS) and two-mode squeezing (TMS) experiments at the two selected flux points. 
Operationally, we drive the device with a CW pump tone at frequency $f_p$ and fully reconstruct the quantum state generated at the TWPA output. 
To do so, we perform repeated acquisitions of the real and imaginary part of the output field quadratures at the frequencies of interest, that is $f_s=(f_p/2-\Delta)$ and $f_i=(f_p/2+\Delta)$, with $\Delta = 0$ for the SMS experiment, and $\Delta \neq 0$ for the TMS experiment. For each experiment, we acquire $N_{\mathrm{rep}}\sim 10^6$ repeated acquisitions, each with an integration time of $10 \,\mu$s. We repeat the same experimental procedure for both pump ON and pump OFF. 

We stress that as the field quadratures are measured at room temperature, the estimation of their values at the output of the TWPA requires normalization by the system gain, $G_{\mathrm{sys}}$. We estimate $G_{\mathrm{sys}}$ by using a shot noise tunnel junction (SNTJ) as a reference noise source in a separated cooldown \cite{malnou_three-wave_2021} (see Appendix \ref{cal_q}).

For Gaussian states, the covariance matrix encodes all the quantum properties of the state under investigation \cite{adesso_continuous_2014}. We estimate the covariance matrix of the quantum state at the output of the TWPA from the normalized experimental quadrature data. 
For the single mode case, we compute the covariance matrix for both pump OFF and pump ON as follows,
\begin{equation}
\sigma^{\text{meas}} = 4
\begin{pmatrix}
\langle \hat{x}^2 \rangle - \langle \hat{x} \rangle^2 & \frac{1}{2} \langle \hat{x}\hat{p} + \hat{p}\hat{x} \rangle - \langle \hat{x} \rangle \langle \hat{p} \rangle \\
\frac{1}{2} \langle \hat{x}\hat{p} + \hat{p}\hat{x} \rangle - \langle \hat{x} \rangle \langle \hat{p} \rangle & \langle \hat{p}^2 \rangle - \langle \hat{p} \rangle^2
\end{pmatrix},
\label{eq:cov_matrix_generic}
\end{equation}
where $\hat{x}=(\hat{a} + \hat{a}^{\dagger})/2$ and $\hat{p}= (\hat{a} - \hat{a}^{\dagger})/2i$ are the normalized single mode field quadratures operators, with $\hat{a}$ and $\hat{a}^{\dagger}$ the bosonic annihilation and creation operators such that [$\hat{a},\hat{a}^{\dagger}$]=1. Finally, we infer the covariance matrix of the quantum state $\ket{\Psi}$ generated at the output of the TWPA by subtracting the pump OFF noise background as follows \cite{flurin_superconducting_2015}
\begin{equation}
\label{sigma_meas}
   \sigma^{\ket{\Psi}}=\sigma^{\text{meas,ON}}-\sigma^{\text{meas,OFF}}+
\mathbbm{1} \, .
\end{equation}
Note that we adopt a convention for field quadratures definition such that the covariance matrix for the vacuum state corresponds to the unit matrix. 

Fig. \ref{fig:SMS_vs_pump_phase}(e) and (g) report examples of the differential (pump ON - pump OFF) phase-space histogram plot of single-mode quadrature experimental data for $N_{\mathrm{rep}} = 8 \times 10^{6}$ repeated acquisitions for the operating flux $\Phi_1$ and $\Phi_2$ respectively, along with the corresponding inferred covariance matrix, Fig. \ref{fig:SMS_vs_pump_phase}(f) and (h). 
\begin{figure*}[htbp]
\includegraphics[width=0.8\linewidth]{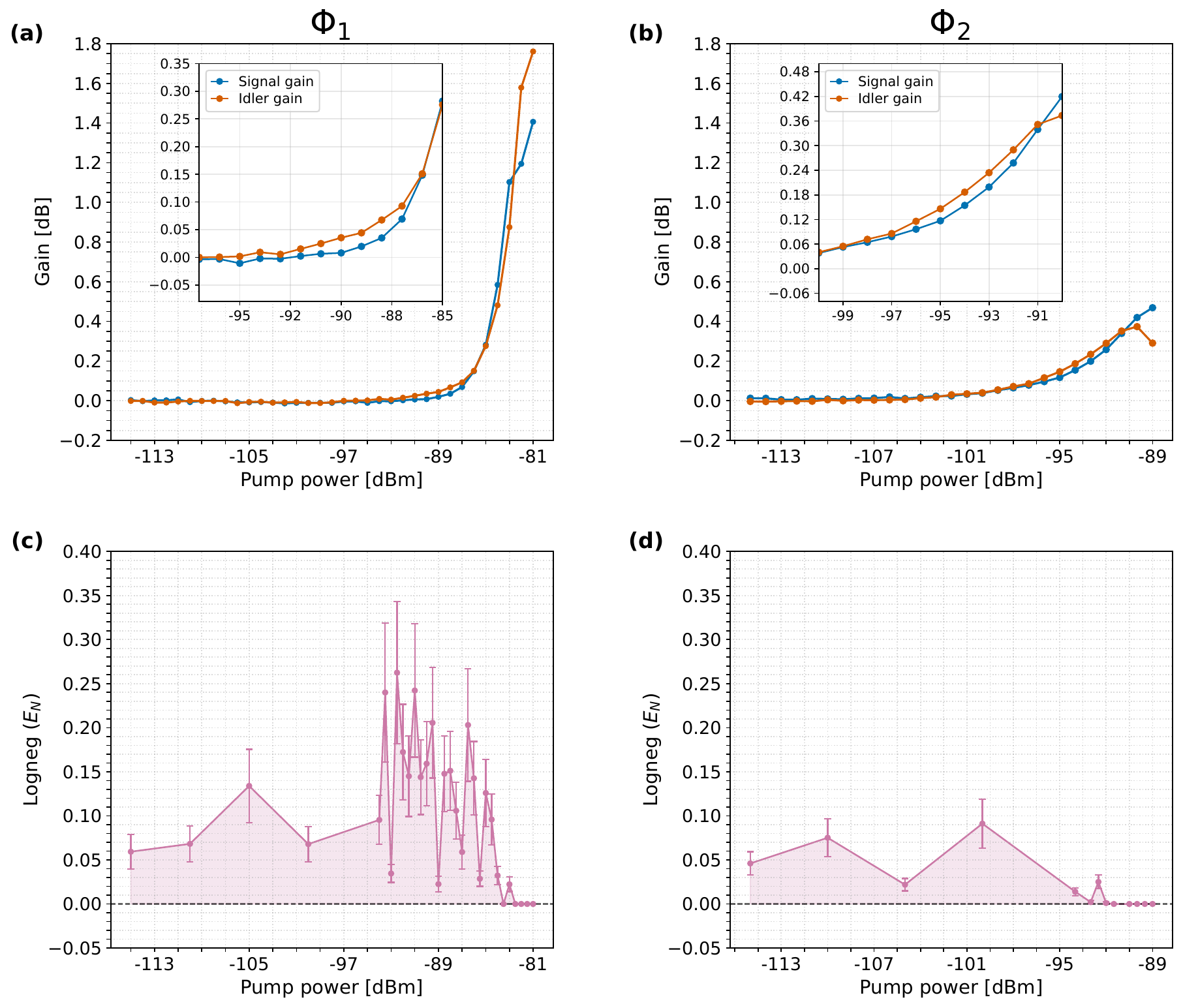}
\caption{\label{fig:TMS_vs_pump_power}
    \textbf{Non-degenerate gain and entanglement verification in two-mode squeezing for $\Phi_1$ (left) and $\Phi_2$ (right).} \textbf{(a-b)} 3WM signal and idler gain experimental results versus pump power at TWPA input; the insets are zoom in of the results; pump frequency $f_p = 7.705~\mathrm{GHz}$; detuning $\Delta= 31 \,\mathrm{MHz}$ for $\Phi_1$ and $\Delta= 61 \,\mathrm{MHz}$ for $\Phi_2$. \textbf{(c-d)} Logarithmic negativity as a function of pump power; number of repeated acquisitions  $N_{\mathrm{rep}} = 1 \times 10^{6}$, integration time for each acquisition $10 \, \mu s$.
    }
\end{figure*}
This experiment is repeated for different values of the input pump phase and the amount of squeezing, $S_{x}$ and $S_{p}$, along the $x$ and $p$ quadratures, is estimated (see Appendix \ref{cov_mat} details on how squeezing is computed from the covariance matrix). The results are shown in Fig. \ref{fig:SMS_vs_pump_phase} (c) and (d) as a function of pump phase for both flux points. 
%
The pump powers are selected in order to have comparable maximum gain for the two operating flux points.

\begin{figure*}[htbp]
    \centering
    \includegraphics[width=0.85\linewidth]{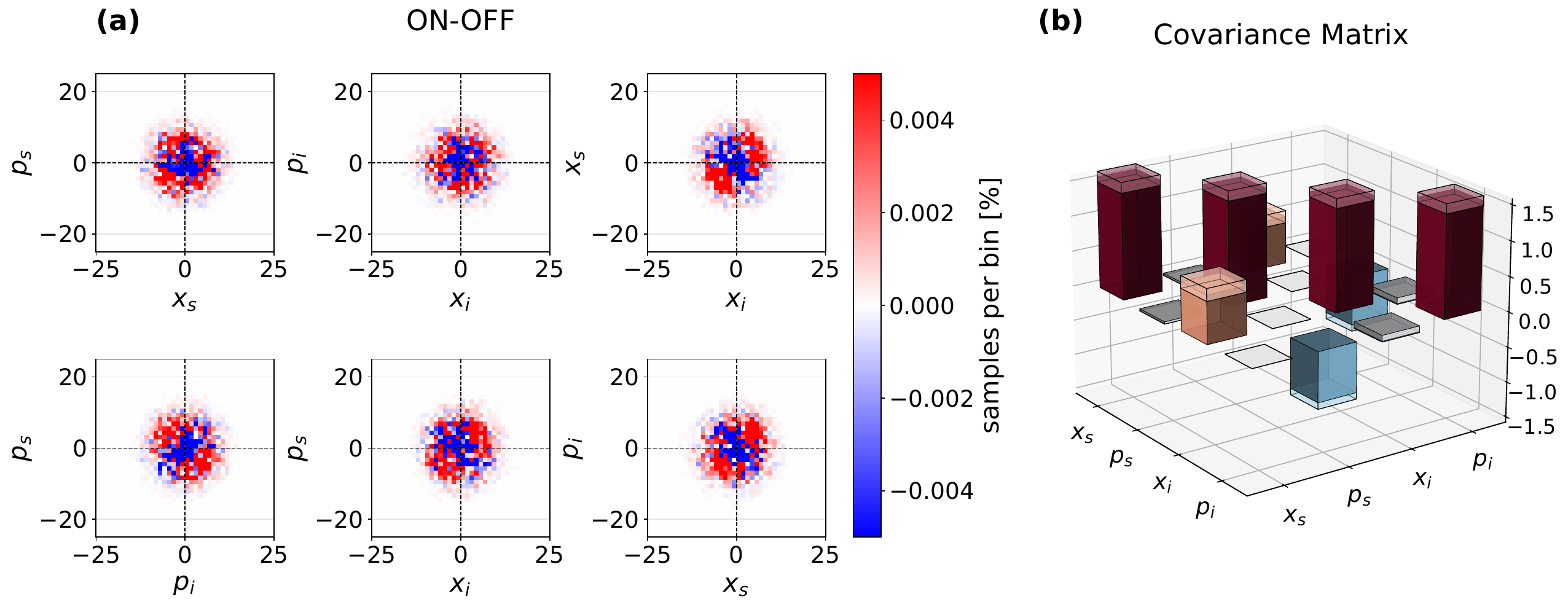}
    \caption{\textbf{Examples of two mode squeezing experimental results.} \textbf{(a)} Differential (pump ON - pump OFF) phase space histogram plots for $N_{\mathrm{rep}} = 8 \times 10^{6}$  acquisitions with integration time $10 \, \mu$s. Flux operating point $\Phi_1$, pump frequency $f_p = 7.705~\mathrm{GHz}$, detuning $\Delta = 31 \,\mathrm{MHz}$, pump power at TWPA input $P_p = -83 \,\mathrm{dBm}$. \textbf{(b)} Corresponding inferred covariance matrix with uncertainty signified by shaded region.}
    \label{fig:TMS_hs}
\end{figure*}

For the optimal flux point $\Phi_1$, our results demonstrate phase-dependent 3WM single-mode squeezing reaching maximum values of about 3 dB below the vacuum level. 
In contrast, for the flux point $\Phi_2$, no squeezing modulation is observed as a function of pump phase, proving the critical role of the operating flux choice for optimizing 3WM single-mode squeezing performance.  

The results of the TMS experiments are reported in Fig. \ref{fig:TMS_vs_pump_power} and \ref{fig:TMS_hs}. 
For the TMS case, we pump the device with a CW tone at $f_p = 7.705~\mathrm{GHz}$ and perform $N_{\mathrm{rep}}$ repeated measurements of the real and imaginary part of the field quadratures at signal and idler frequency $f_s = f_p/2 - \Delta$ and  $f_i = f_p/2 + \Delta$. We repeat the experiment for pump ON and pump OFF and from the normalized quadrature data we compute the covariance matrix as follows,
\begin{equation}
    \sigma^{\text{meas}}_{mn}
    = 4\left[
        \frac{1}{2}\langle R_m R_n + R_n R_m \rangle
        - \langle R_m \rangle \langle R_n \rangle
      \right],
    \label{eq:cov_definition}
\end{equation}
with \( R = (\hat{x}_s,\, \hat{p}_s,\, \hat{x}_i,\, \hat{p}_i)\) the array of field quadratures operators including both signal and idler modes (see Appendix \ref{cov_mat}). Finally, we infer the bipartite covariance matrix using Eq. \eqref{sigma_meas}. 

The experiments and the associated estimation of the covariance matrix are repeated for different values of the input pump power. For each pump power we estimate the logarithmic negativity, defined as  $E_N = \max\!\left[-\ln(\nu_-),\,0\right]$, where $\nu_-$ is the smallest symplectic eigenvalue of the partially transposed covariance matrix (see Appendix \ref{cov_mat}). 
According to the Positive Partial Transpose (PPT) criterion \cite{horodeckiSeparabilityCriterionInseparable1997}, the condition $E_N>0$ is necessary and sufficient to demonstrate entanglement between the signal and idler modes.

The results for the TMS experiments versus pump power are reported in Fig. \ref{fig:TMS_vs_pump_power} for both operating flux points. Panels (a-b) show the signal and idler gain, while panels (c-d) report the estimated logarithmic negativity for the bipartite state at the TWPA output. The detuning $\Delta = (f_i -f_s)/2$ 
 is chosen to ensure symmetric gain for signal and idler modes \cite{houde_loss_2019}.

In the low pump power regime (no TWPA gain), the logarithmic negativity is expected to be zero, corresponding to uncorrelated quadratures in the two-mode vacuum state. 
 The observed non-zero values of $E_N$ at low pump powers for both flux biases can be attributed to experimental artifacts, reported also in similar experimental demonstrations \cite{Perelshtein22}, such as drifts in the amplification chain occurring between the measurement sequences \cite{jarvis-frain_observation_2025}.


For the optimal flux point $\Phi_1$, as the pump power increases, we observe an increase of the logarithmic negativity with respect to the values in the low pump power regime, evidencing two-mode entanglement generation, followed by a decrease to zero values for higher pump powers. 
Fig. \ref{fig:TMS_hs} shows an example of differential (pump ON - pump OFF) histogram plots of the two-mode experimental quadrature data and the corresponding reconstructed covariance matrix for the optimal flux point $\Phi_1$. 

From results in Fig. \ref{fig:TMS_vs_pump_power}(c), it is interesting to notice that the logarithmic negativity $E_N$ approaches zero at high pump powers while the gain continues to increase. A similar behavior has been previously observed in TWPA squeezers \cite{Perelshtein22, Esposito22}, suggesting the onset of pump-power dependent mechanisms responsible for entanglement degradation. A detailed study of this behavior is out of the scope of this work.

Finally, in contrast to the flux point $\Phi_1$ where our results indicate two mode entanglement generation, for the flux point $\Phi_2$ (results shown in Fig. \ref{fig:TMS_vs_pump_power}(d)), the logarithmic negativity doesn’t show any clear increase in the pump power range corresponding to similar gain values obtained for $\Phi_1$, denoting again a detrimental role of competing 4WM processes on squeezing. Thus, our findings indicate that a preliminary optimization of the external magnetic flux working point aiming at minimizing the generation of 4WM idler is necessary in order to optimize 3WM single and two mode squeezing performance in flux tunable JTWPA.

\section{Conclusions}
This work presents the first experimental demonstration of squeezing generation via residual 3WM in a SNAIL JTWPA  with alternated flux polarity. 
Operating in the 3WM regime has the advantage of shifting the signal/idler band away from the pump frequency, allowing easy filtering of the pump and thus simplifying applications. 

In addition to showing the versatility of Josephson metamaterials as TWPA squeezers, our work reports a detailed investigation on the impact of simultaneous activation of 3WM and 4WM nonlinearities on the squeezing performance.

We systematically investigate the residual 3WM regime versus the external applied magnetic flux and find that 3WM single and two-mode squeezing can be achieved provided that 4WM nonlinearity is carefully mitigated by an appropriate choice of the operating flux point. 

Our results identify flux-bias control and competition between nonlinear mixing processes as key factors for squeezing performance, highlighting the need for optimization of these aspects in the design and operation of JTWPA squeezers. 

\begin{acknowledgments}
This project has received funding from the European Union under Horizon Europe 2021–2027 Framework Programme Grant Agreement No. 101080152, project TruePA, and under Horizon 2020 Research and Innovation Programme Grant Agreement No. 101017733, project “Superconducting quantum-classical linked computing systems (SuperLink)”, in the frame of QuantERA2 ERANET COFUND in Quantum Technologies"; and by Next Generation EU, Mission 4, NQSTI CUP B53C22004180005. \\
The authors thank Paolo Scotto Di Vettimo, CNR SPIN Naples, for his extensive technical assistance in setup preparation. The authors also thank Nicolas Roch and Luca Planat for insightful comments, and the QTLab group at University of Naples Federico II for helpful discussion. Finally, the authors thank Florent Lecocq and Jos\'e Aumentado for providing the SNTJ and for useful comments.
\end{acknowledgments}

\section*{DATA AVAILABILITY}
The data that support the findings of this article are openly available on Zenodo id https://doi.org/10.5281/zenodo.19701770.
\appendix
\section{Device parameters and SNAIL unit cell}
\label{beta_gamma}
The adopted device was fabricated at the Nanofab facility of the Institut Néel, CNRS, in Grenoble. The fabrication process is nominally identical to the one described in \cite{Ranadive22}, and it is based on a top-ground microstrip aluminum transmission line and Al-AlOx-Al Josephson junctions. 
The main circuit parameters, extracted from experimental characterizations \cite{Ranadive22}, are listed in Table \ref{tab:700_params}.
\begin{table}[h]
\centering
\begin{tabular}{ll}
\toprule
Number of unit cells & 700 \\
\midrule
Josephson capacitance, $C_J$ & $50\,\mathrm{fF}$ \\
\midrule
Ground capacitance, $C_g$ & $250\,\mathrm{fF}$ \\
\midrule
Critical current, $I_c$ & $2.19\,\mu\mathrm{A}$ \\
\midrule
SNAIL ratio, $r$ & 0.07 \\
\bottomrule
\end{tabular}
\caption{Circuit parameters for the adopted device \cite{Ranadive22}.}
\label{tab:700_params}
\end{table}

The unit cells of the adopted JTWPA contain one SNAIL each, with alternated magnetic flux polarity throughout the chain. For a single SNAIL, considering the phase drop across the large Josephson junction as $\phi_L$ and across the small junction as $\phi$, and by using the flux quantization for a single loop, we can write the difference between the phase drops across the lower and upper branches of the superconducting loop as follows, 
\begin{equation}
    \phi -3\phi_L =\phi_{\text{ext}},
    \label{phase_loop}
\end{equation}
where $\phi_{\text{ext}}$=$2\pi \Phi_{\text{ext}}/\Phi_0$ is the reduced external magnetic flux, with $\Phi_0$=$h/(2e)$ the magnetic flux quantum. By using equation \eqref{phase_loop}, the current through each SNAIL element can be expressed as
\begin{equation}
    I=rI_c\sin{\phi}+I_c\sin\left({\frac{\phi -\phi_{ext}}{3}}\right).
    \label{curr_phase}
\end{equation}
By performing a Taylor expansion of $I$ about $\phi^*$ such that $I(\phi^*)=0$, we get the following approximated expression
\begin{equation}
\frac{I(\phi^{*} + \phi)}{I_c}
\approx
\left.\frac{\partial I}{\partial \phi}\right|_{\phi^{*}} \phi
+ \frac{1}{2}\left.\frac{\partial^2 I}{\partial \phi^2}\right|_{\phi^{*}} \phi^2
+ \frac{1}{6}\left.\frac{\partial^3 I}{\partial \phi^3}\right|_{\phi^{*}} \phi^3 \quad ,
\label{eq:taylor}
\end{equation}
that can be rewritten as in Equation \eqref{curr_phi} in the main text.

\section{Experimental setup}
\begin{figure*}[htbp]
    \centering
    \includegraphics[width=0.8\linewidth]{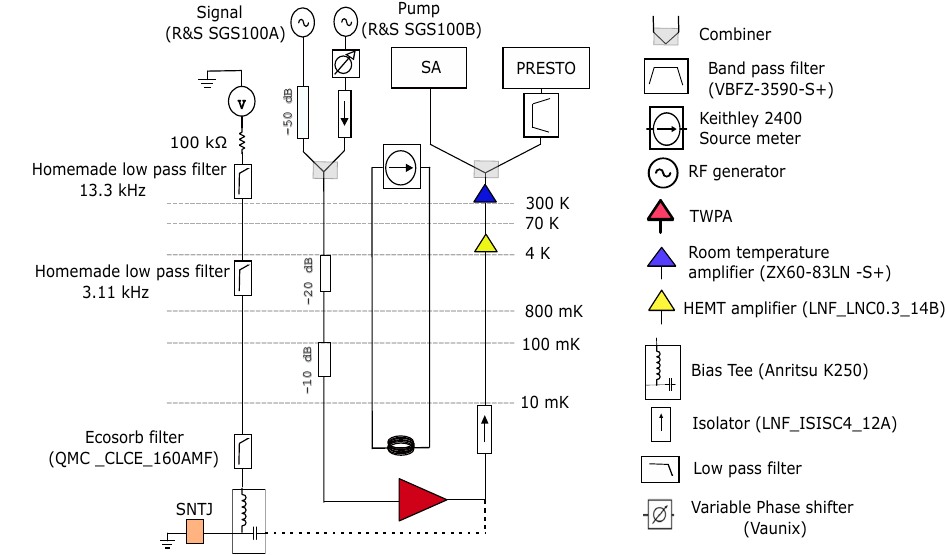}
    \caption{\textbf{Sketch of the experimental setup.} The dashed line indicates the setup used in the cooldown dedicated to the system gain calibration in which the TWPA device is substituted with a shot noise tunnel junction (SNTJ).}
    \label{fig:setup}
\end{figure*}

A schematic of the experimental setup is shown in Fig. \ref{fig:setup}. 
\label{exp_setup}
The sample is anchored to the mixing chamber plate of a dilution refrigerator ($Triton^{TM} 400$ from Oxford Instruments). The external magnetic flux is controlled by injecting DC current in a superconducting coil placed alongside the device. 

Two different cryogenic setups, in two separated cooldowns, have been adopted: one, including the TWPA, for the gain and squeezing experiments presented in the main text, and a second one, excluding the TWPA, for calibrating the system gain.

For the system gain calibration cooldown, the TWPA device is substituted with a shot noise tunnel junction (SNTJ) which is used as calibrated noise source. The adopted SNTJ has been fabricated by the team of José Aumentado and Florent Lecocq at NIST, Boulder, Colorado, USA \cite{malnou_low-noise_2024}. 
A voltage bias can be applied to the SNTJ through a DC input line with low pass filter stages at room temperature, at the 4 K plate and at the mixing chamber plate. 

At room temperature, two CW RF sources (R\&S SGS100A and SGS100B) are used for generating the input pump tone at frequency $f_p$ and signal tone at frequency $f_s$. A tunable phase shifter allows to control the pump phase. 


For the experimental results reported in Fig. \ref{fig:flux_sweet_spot} at room temperature we measure the idler power spectral density (PSD) in dBm with a spectrum analyzer with set resolution bandwidth $3 \, \text{kHz}$. 

For the squeezing experiments, the output field quadratures are acquired using a room temperature microwave platform called Presto, from Intermodulation Products AB \cite{Tholen:2022xpq}. Such acquisition system does not require the use of analog mixers for down conversion and it easily enables the simultaneous detection of multiple frequency tones by assigning a single local oscillator (LO) and different intermediate frequencies (IFs). 
In our case, we operate the Presto in lock-in mode, with local oscillator frequency (LO) set to be  $f_p/2$ and a single intermediate frequency (IF) equal to $\Delta$. The lockin measurement bandwidth is set to $100 \, \mathrm{kHz}$ in order to get  $N_{\mathrm{rep}}$ repeated field quadrature measurements, each with $10 \, \mu$s integration time. For the degenerate gain measurements and single mode squeezing acquisitions, we set $\Delta = 0$ in order to acquire exactly at half of the pump frequency. For the non-degenerate gain and two mode squeezing, we set $\Delta \neq 0$ and acquire the field quadratures at the upper sideband (USB) and lower sideband (LSB) corresponding to our idler and signal frequencies.
%
\section{WRSPICE Numerical simulations}
\label{simulation}
The adopted JTWPA has been modeled at the circuit level using WRspice. A complete description of the device model adopted in simulations, including the netlist, is available at \cite{levochkina_github_2024}. 

To reproduce the frequency and flux dependence of the 3WM idler observed in experiments, it is essential to include realistic imperfections in the JTWPA model. We model these imperfections by introducing variations in the Josephson junctions’ critical currents, assigning each junction a value that deviates by up to ±5$\%$ from the nominal value, consistent with the analysis reported in \cite{levochkina_investigating_2024}.

TWPA dielectric losses are also modeled in the simulations, by using equivalent series resistors for each capacitor $C_g$, by considering the dielectric loss tangent estimated from experimental characterizations \cite{Ranadive22, ranadive_nonlinear_2022}. 
The simulated output signal reaches a steady state at 4.406 ns
\cite{levochkina_numerical_2024}; to ensure stability, we analyze the simulated data  after 10 ns. A 60 ns measurement window is used, yielding a frequency resolution of approximately 16.7 MHz for the Fourier transform of the raw time-domain signal.
The TWPA was driven in simulation with a pump and a signal tones (peak currents of $0.157 \mu$A and $0.0011 \mu$A, respectively), ensuring operation well below saturation in both the 3WM and 4WM regimes.

The results of the simulations are reported in Fig. \ref{fig:flux_sweet_spot}(c) and compared with the experimental results. We stress that no fitting parameters are used in the simulations.


\section{Quadrature normalization procedure}
\label{cal_q}
To compare the measured field quadrature variances with the vacuum noise level, the quadratures values acquired with the Presto electronics at room temperature must be normalized in order to be referred to the output of the TWPA. We convert the quadratures $X_{FS}$ and $P_{FS}$, provided by the Presto electronics in full scale (FS) units into normalized quadratures $X$ and $P$, in unit of square root photon number, by multiplying them with a normalization factor $\upsilon$:

\begin{equation}
    X= \upsilon\, X_{\text{FS}}, 
    \qquad
    P = \upsilon\, P_{\text{FS}} \, .
\end{equation}\\
The normalization factor \cite{Perelshtein22} is given by
\begin{equation}
    \upsilon = \varepsilon \sqrt{\frac{\eta\, t_{\mathrm{int}}}{G_{\mathrm{sys}}Z_0\, h\, f_{\mathrm{acq}}}},
\end{equation}
where $\eta$ indicates the internal loss of the TWPA device, $G_{\mathrm{sys}}$ is the estimated system gain, $Z_0 = 50\,\Omega$ is the characteristic impedance of the microwave lines, $h$ is Planck's constant, $f_{\mathrm{acq}}$ is the acquisition frequency, $t_{\mathrm{int}}$ is the integration time window, and $\varepsilon = 0.98$ is the calibrated Presto conversion coefficient which converts full scale (FS) units into Volts units. \\
Losses in the adopted JTWPA device are primarily arising from the dielectric losses in the capacitance to ground per unit cell, which is required for impedance matching. The insertion loss of the JTWPA is calculated from the dielectric loss tangent \cite{planat_fabrication_2019}, $\tan\delta_0 = 2.1 \times 10^{-3}$, estimated from experimental characterizations \cite{Ranadive22, ranadive_nonlinear_2022}.

To estimate the system gain $G_{\text{sys}}$, we perform a separated cooldown in which we substitute the TWPA device with a shot noise tunnel junction (SNTJ) leaving the rest of the setup identical (see Fig. \ref{fig:setup}). 

The noise power emitted by the SNTJ in the quantum regime ($k_BT<<hf$) at frequency $f$ as a function of the bias voltage $V$ can be expressed as
\begin{equation}
\label{noise_SNTJ}
\begin{aligned}
&N_{\mathrm{SNTJ}}(f,V) = \Bigg[\frac{1}{2} \Bigg[
\frac{eV+hf}{2k_B}
\coth\!\left(\frac{eV+hf}{2k_B T}\right) \\
+\, 
&\frac{eV-hf}{2k_B}
\coth\!\left(\frac{eV-hf}{2k_B T}\right)
\Bigg] +T_{\text{sys}} \Bigg] \, \text{BW} \, G_{\text{sys}} \, k_B \, ,
\end{aligned}
\end{equation}
where $\text{BW}$ denotes the measurement bandwidth.

\begin{figure}[htbp]
    \centering
    \includegraphics[width=1\linewidth]{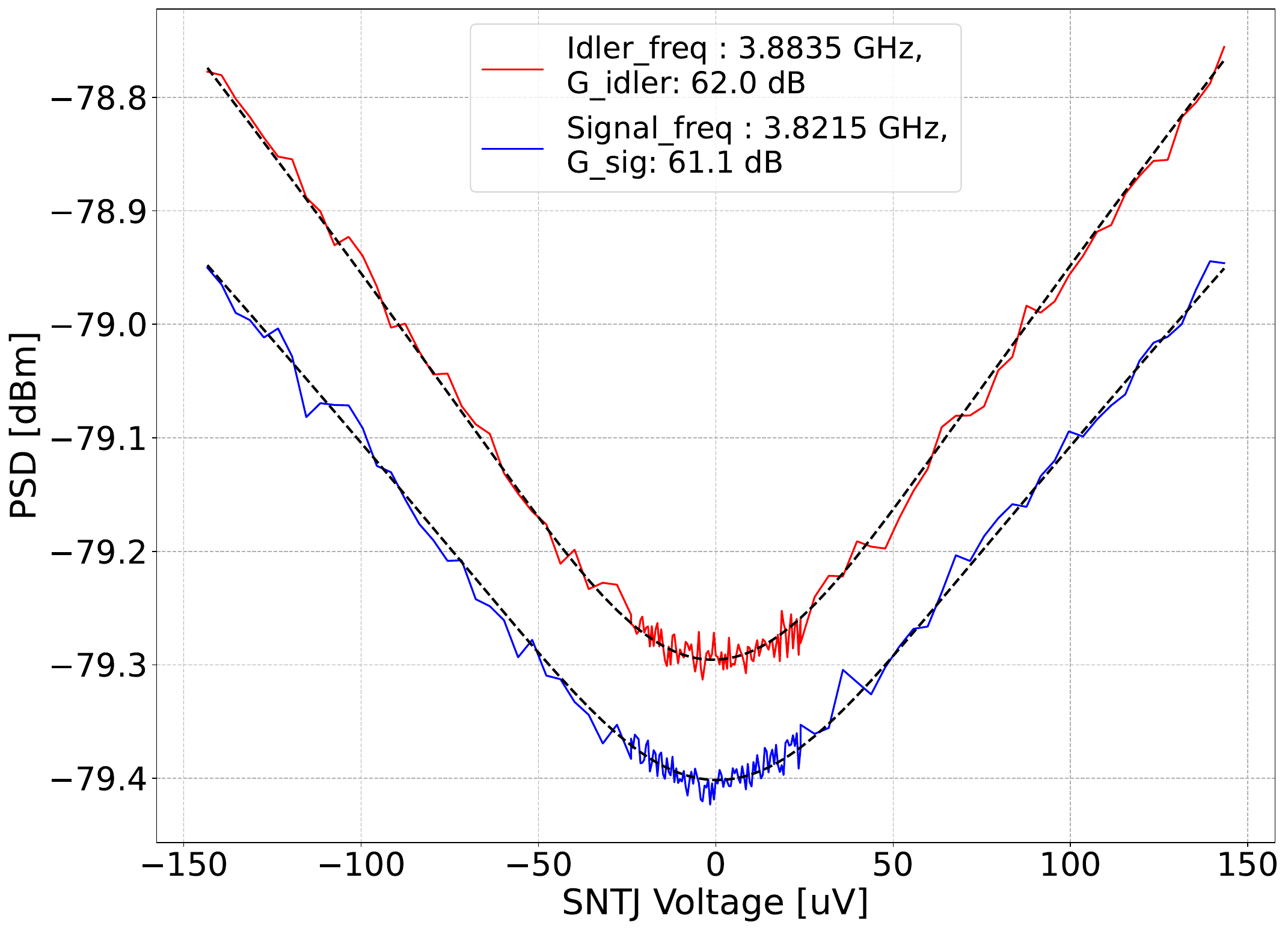}
    \caption{\textbf{$G_{\text{sys}}$ estimation using SNTJ.} Example of measured PSD as a function of SNTJ voltage bias for a pair of signal and idler frequencies. Best fits are shown with dashed lines. $G_{\text{sys}}$ best fit values are reported in the legend.}
    \label{fig:system_gain}
\end{figure}

We measure the power spectral density (PSD) recorded by a spectrum analyzer as a function of the bias voltage $V$ applied to the SNTJ. We repeat this measurement for each of the frequencies of interest. For biasing, we use a filtered DC line with a 100 k$\Omega$  resistance at room temperature, which works as a voltage divider avoiding accidental blow-up \cite{malnou_three-wave_2021}. We fit the obtained PSD to the noise model in Equation \eqref{noise_SNTJ} using three free parameters, the system gain, $G_{\text{sys}}$, the system noise temperature $T_{\text{sys}}$ and the SNTJ electronic temperature $T$. Fig. \ref{fig:system_gain} shows an example of measured power spectral density (PSD) as a function of the SNTJ bias voltage, along with the corresponding best-fit curve for a pair of frequencies signal/idler frequency tones. 

We stress that the system gain obtained from the above method corresponds to the gain from the reference plane of the SNTJ to the input of the spectrum analyzer. The estimated system gains at the frequencies of interest are listed in Table \ref{tab:sys_gain}.
\begin{table}[h]
\centering
\begin{tabular}{ll}
\hline
Acquisition frequency & $G_{\text{sys}}$ [$\mathrm{dB}$] \\
\hline
$f_p/2$ & 61.7 \\
\hline
$f_p/2+31\,\mathrm{MHz}$ & 62.0  \\
\hline
$f_p/2-31\,\mathrm{MHz}$ & 61.1   \\
\hline
$f_p/2+61\,\mathrm{MHz}$ & 61.5   \\
\hline
$f_p/2-61\,\mathrm{MHz}$ & 62.0 \\
\hline
$f_p + 31\,\mathrm{MHz}$ & 46.5 \\
\hline
\end{tabular}
\caption{Calibrated system gain for the different acquisition frequencies. $f_p = 7.705~\mathrm{GHz}$. }
\label{tab:sys_gain}
\end{table}
We estimate an upper bound of 1  dB for the total insertion loss between the SNTJ output reference plane and the cryogenic isolator input reference plane, including loss from the SNTJ packaging and the used bias tee \cite{chang_noise_2016}. The $G_{\text{sys}}$ value obtained from the SNTJ calibration is corrected by such upper bound loss estimation before it is used for the normalization of the quadrature data. This means that we use an upper bound estimation of $G_{\text{sys}}$ and consequently a lower bound (conservative) estimation for squeezing and logarithmic negativity.\\

Considering the typical ripples in the insertion loss characterization of the microwave components between SNTJ and the isolator at mK temperature, we assume an uncertainty of 1 dB in our final estimation of $G_{\text{sys}}$. The error bars on the estimated covariance matrix and consequently on the estimation of squeezing and logarithmic negativity in this work are obtained assuming that the error on the $G_{\text{sys}}$ estimation is the dominant one.

\section{Estimation of the total input attenuation}

\label{att}

The input pump and signal power reported in this work are all refereed at the TWPA input. To do so we sum the power at the output of our RF source at room temperature to the total input line attenuation $A_{\text{in}}$, which indicates attenuation from the room temperature microwave sources to the input of the device at mK temperature. The latter is estimated using the following formula, 
\begin{equation}
    S_{21}^{\text{off}} = A_{\text{in}}+\eta+G_{sys},
\end{equation}
where $S_{21}^{\text{off}}$ is the transmission measured through the JTWPA when the pump is off, $G_{sys}$ is the calibrated system gain, and $\eta$ is the insertion loss of the TWPA in dB, estimated from the loss tangent, $\tan\delta_0 = 2.1 \times 10^{-3}$, obtained in experimental characterizations \cite{Ranadive22, ranadive_nonlinear_2022}


\section{Squeezing and Logarithmic negativity}
\label{cov_mat}
The amount of single mode squeezing in dB along the $x$ and $p$ quadratures can be defined starting from the estimated single mode covariance matrix $\sigma^{\ket{\Psi}}$ (Eq. \eqref{sigma_meas}) as follows,
\begin{equation}
    S_{x} = 10 \log_{10} \left( \frac{\sigma^{\ket{\Psi}}_{11}}{\sigma_{11}^{|0\rangle}} \right),
\qquad
    S_{p} = 10 \log_{10} \left( \frac{\sigma^{\ket{\Psi}}_{22}}{\sigma_{22}^{|0\rangle}} \right) \, .
\end{equation}
With our convention for the field quadrature definition, the covariance matrix of the single mode vacuum state $\sigma^{|0\rangle}$ correspond to the unity matrix. We use the definitions above to get the results in Fig. \ref{fig:SMS_vs_pump_phase}(c-d).

For the TMS, the main quantity of interest for the entanglement verification is the logarithmic negativity defined as,
\begin{equation}
    E_N = \max\!\left[-\ln(\nu_-),\,0\right],
    \label{eq:EN_def}
\end{equation}
where $\nu_-$ is the smallest symplectic eigenvalue of the partially transposed two-mode covariance matrix. This eigenvalue is given by,
\begin{equation}
\begin{aligned}
    \nu_- =
    \sqrt{
        \frac{
            \Delta\sigma - \sqrt{(\Delta\sigma)^2 - 4\det\sigma}
        }{2}
    },\\
    \Delta\sigma = \det A + \det B - 2\det C,
    \label{eq:nu_minus}
\end{aligned}
\end{equation}
with $A$, $B$, and $C$ denoting the submatrices of the two mode covariance matrix expressed in the following block structure,
\begin{equation}
    \sigma =
    \begin{pmatrix}
        A & C \\[4pt]
        C^T & B
    \end{pmatrix}.
    \label{eq:cov_block_form}
\end{equation}
According to the Positive Partial Transpose (PPT) criterion~\cite{simon_peres-horodecki_2000, adesso_gaussian_2005}, a bipartite Gaussian state is separable only if the smallest symplectic eigenvalue of the partially transposed covariance matrix satisfies $\nu_- \ge 1$. Consequently, a sufficient condition for entanglement, is $\nu_- < 1$ or equivalently $E_N > 0$.


\bibliography{Bibliography}

\end{document}